\documentclass[useAMS,usenatbib,twocolumn]{aastex62}
\usepackage{times,graphicx,amsmath,amsfonts,amssymb}
\usepackage{epsfig}

\usepackage[normalem]{ulem} % either use this (simple) or
\usepackage{tabularx}

\usepackage{array}
\newcolumntype{L}[1]{>{\raggedright\let\newline\\\arraybackslash\hspace{0pt}}m{#1}}
\newcolumntype{C}[1]{>{\centering\let\newline\\\arraybackslash\hspace{0pt}}m{#1}}
\newcolumntype{R}[1]{>{\raggedleft\let\newline\\\arraybackslash\hspace{0pt}}m{#1}}

\newcommand{\hi}{\mbox{H{\scriptsize I}}}

\def\ltsim{\lower.5ex\hbox{$\; \buildrel < \over \sim \;$}}
\def\gtsim{\lower.5ex\hbox{$\; \buildrel > \over \sim \;$}}
\def\ltsim{\lower.5ex\hbox{$\; \buildrel < \over \sim \;$}}
\def\gtsim{\lower.5ex\hbox{$\; \buildrel > \over \sim \;$}}
\def\be{\begin{equation}}
\def\ee{\end{equation}}
\def\ba{\begin{eqnarray}}
\def\ea{\end{eqnarray}} 

\def\kms{\, {\rm km }\, {\rm s}^{-1}}
\def\mnras{MNRAS}

\def\ln{{\rm ln}\,}

\def\frac#1#2{{\textstyle{#1\over #2}}}

\newcommand{\NDF}{\textsc{NGC1052-DF2}}

\shorttitle{Dynamical friction of Globular Clusters in \NDF\ }
\shortauthors{A. Nusser}
\begin{document}

\title{Orbital decay of Globular Clusters in the galaxy with little dark matter}
\email{adi@physics.technion.ac.il} 
\author{Adi Nusser}
\affiliation{Department of Physics and the Asher Space Research Institute, Israel Institute of Technology Technion, Haifa 32000, Israel}

%\date{Accepted XXXXXX. Received XXXXXX; in original form XXXXXX}
%\pagerange{\pageref{firstpage}--\pageref{lastpage}} \pubyear{2018}

\begin{abstract}

Recently, \cite{vanDokkum2018} have presented an important discovery of an  ultra-diffuse galaxy, \NDF ,  with a dark matter content significantly less than predicted from its  stellar mass alone. The analysis relies on 
 measured radial velocities of 10  Globular Clusters (GCs), of estimated individual masses 
 of a few $ \times  10^6 M_\odot$. 
 This is about $1\%$ of the inferred mass of \NDF\  of  $2\times 10^8 M_\odot$ within  a half-light radius, 
$R_\mathrm{e}=2.2\, \mathrm{kpc}$. The  large relative mass and the old age of these objects imply that they might be susceptible to orbital decay by dynamical friction.   Using analytic estimates and N-body simulations  of an isolated system matching the inferred mass profile of \NDF , we show that orbits of the most massive GCs  should already have decayed  on a time scale of a few Gyrs.
These findings should help in constraining  mass profile and formation scenarios of \NDF .

\end{abstract}


\section{Introduction}
\label{sec:intro}

Small galaxies with shallow potential wells are an interesting probe of the nature of dark matter (DM). 
The sheer existence of DM in these objects has long ago allowed a constraint on the mass of neutrinos as candidates for DM \citep{Tremaine1979}. 
Further, their observed abundance  is sensitive to both the power spectrum of the initial mass fluctuations \citep{Lovell2013}  and 
energetic feedback associated with star formation and Supernovae feedback \citep{Pontzen2012}. 
The ratio of DM to stellar mass of these galaxies  is $\gtsim 100$ \citep{Behroozi2010,Moster2013}. Therefore, the discovery  
of a galaxy, \NDF , with  very little or no DM content at all is particularly intriguing \citep{vanDokkum2018}. 
A  DM deficiency in \NDF\ in conjunction with the low acceleration can in principle constrain  modifications to Newtonian dynamics 
\citep[e.g.][]{vanDokkum2018,Famaey2018,Moffat2018}.
A lack of DM could serve as an indication of dark sector breaking of the equivalence principle and violation of Lorenz invariance \citep[e.g.][]{Frieman1991,Kesden2006,Keselman2009,Bettoni2017}. 
Indeed, an additional fifth force in the dark sector could completely segregate the stellar and DM components
due to the gravitational force field of a more massive host, such as the large elliptical NDF1052 in the vicinity of \NDF .
%\label{sec:appendix}

The analysis of \NDF\ is based on the measured radial velocities of ten compact objects which are similar to Galactic Globular Clusters (GCs)  and are hence 
termed so by \cite{vanDokkum2018}. The 90\% confidence limit on the line of sight velocity dispersion of these objects is estimated as 
$\sigma <10.5\kms$ \cite{vanDokkum2018}.
\cite{Martin2018} have argued that the small number of tracers used to constrain the kinematics of the galaxy could be associated with 
poorly determined velocity dispersion and hence is the cause of the apparent lack of DM. 

The GCs in \NDF\ are much more luminous than typical GCs with the brightest of them, GC-73,  having an absolute luminosity and metallicity similar to
$\omega$ Cen \citep{vanDokkum2018a}, the  brightest GC in the Milky Way. We  infer the mass of GC-73 to be $\sim 3.5-4\times 10^6 M_\odot$, the close to  that of 
$\omega$ Cen \citep{Souza2013}. The second brightest  object, GC-77, is consequently only a factor of 1.6 less massive.
 Thus  the mass  one of the  brightest GCs  is  about 1\% of the high end mass estimate of \NDF , as provided by \cite{vanDokkum2018}. This  makes these 
objects particularly susceptible to dynamical friction \citep{Chandrasekhar1949}. The time scale for orbital decay of a GC of mass $M_\mathrm{GC}$ in a galaxy with mass profile $M_\mathrm{gal}(r)$ is \citep{Binney2008}, 
\begin{equation}
\label{eq:DF}
t_\mathrm{DF}=\frac{1.17}{\ln\Lambda} \frac{M_\mathrm{gal}(r)}{M_{\mathrm GC}} t_\mathrm{cross}\; .
\end{equation}
The derivation of this relation employs several  assumptions about the structure of the galaxy which do not necessarily hold in reality  \citep[c.f.][for details]{Binney2008,Arca2016}. Therefore, we use it to infer a rough estimate only.
Consider an object at a $2\, \mathrm{kpc}$ distance from the center and with a velocity of $5 \kms$. Then, $t_\mathrm{cross}\approx 2\, \mathrm{kpc}/5\kms\approx 0.4 \mathrm{Gyr}$. If the mass ratio is $\approx 100$ and $\ln\Lambda \approx6 $, 
we find
$t_\mathrm{DF}\ltsim 8 \mathrm{Gyr}$. Given that the estimated age of the GC is  $ \gtsim 9\mathrm{Gyr}$ \citep{vanDokkum2018a}, orbit decay by DF should be taken into account. 

In the remainder of the Letter, we provide  a more robust assessment of the effects of dynamical friction using N-body simulations designed to model the \NDF\ system.

%%%%%  SIMULATIONS %%%%%%%
\section{Simulations}
\label{sec:simulations}

We study the orbital decay using  N-body simulations of an isolated collisionless system with density profile matching the 
general features  of  \NDF , as reported in \cite{vanDokkum2018}.
%% MASS PROF
We model the galaxy as a two component system of stars and DM, both  assumed spherical with  respective \citep{Einasto89} density profiles. 
The observed  two-dimensional (2D) structure of the galaxy is represented by \cite{vanDokkum2018} in terms of a S\'ersic profile with index $n_S=0.6$ 
and half-light radius $R_e\approx 2.2\, \mathrm{kpc}$. We have found that this is very well approximated as a projection of the 
 three-dimensional (3D) Einasto density profile with   parameters  $r_v=10$ kpc, $r_{-2}=r_v/8.1$ and $n_E=2.5$. The total mass in stars is normalized to $2\times 10^8 M_\odot$, as in the observations. 
 The stellar component is  assumed to be embedded in a DM halo with an Einasto profile with parameters derived from halos identified in large high-resolution cosmological simulations \citep[e.g.][]{Ludlow2013}. We consider halos of virial masses $10^8M_\odot$ and $10^9M_\odot$ corresponding to virial radii 
 $r_v=10 $kpc and $r_v=21.6$ kpc, respectively. For both masses  we fix the Einasto parameters at $r_{-2}=r_v/20$ and $n_E=6$.
  \cite{vanDokkum2018} derive the constraints on the mass profile taking a distance of $20 \, \mathrm{Mpc}$ for \NDF . 
  There is, however, a debate regarding the distance.
 \cite{Trujillo2018} have  presented arguments that the galaxy may  be much nearer at a distance of $D=13\, \mathrm{Mpc}$.   \cite{vanDokkum2018b} countered these arguments, producing a revised distance of $19\pm 1.7\mathrm{Mpc}$. Nonetheless, here we 
  also model GC  orbits for  parameters appropriate for $D=13\, \mathrm{Mpc}$,
where   the spacial extent of the galaxy is reduced to $R_e\approx1.4\, \mathrm{kpc}$ and its stellar mass to $~6\times  10^7 M_\odot$. Further, we model the stellar component is an Einasto profile with the same $r_v$ and $n_E$ as above but with $R_{-2}=r_v/12.1$ and mass normalized to $6\times  10^7 M_\odot$. The parameters of the DM halo remain as before with a virial  mass of $10^9 M_\odot$, consistent with  \cite{Trujillo2018}.

\begin{figure}
 \includegraphics[width=0.48\textwidth]{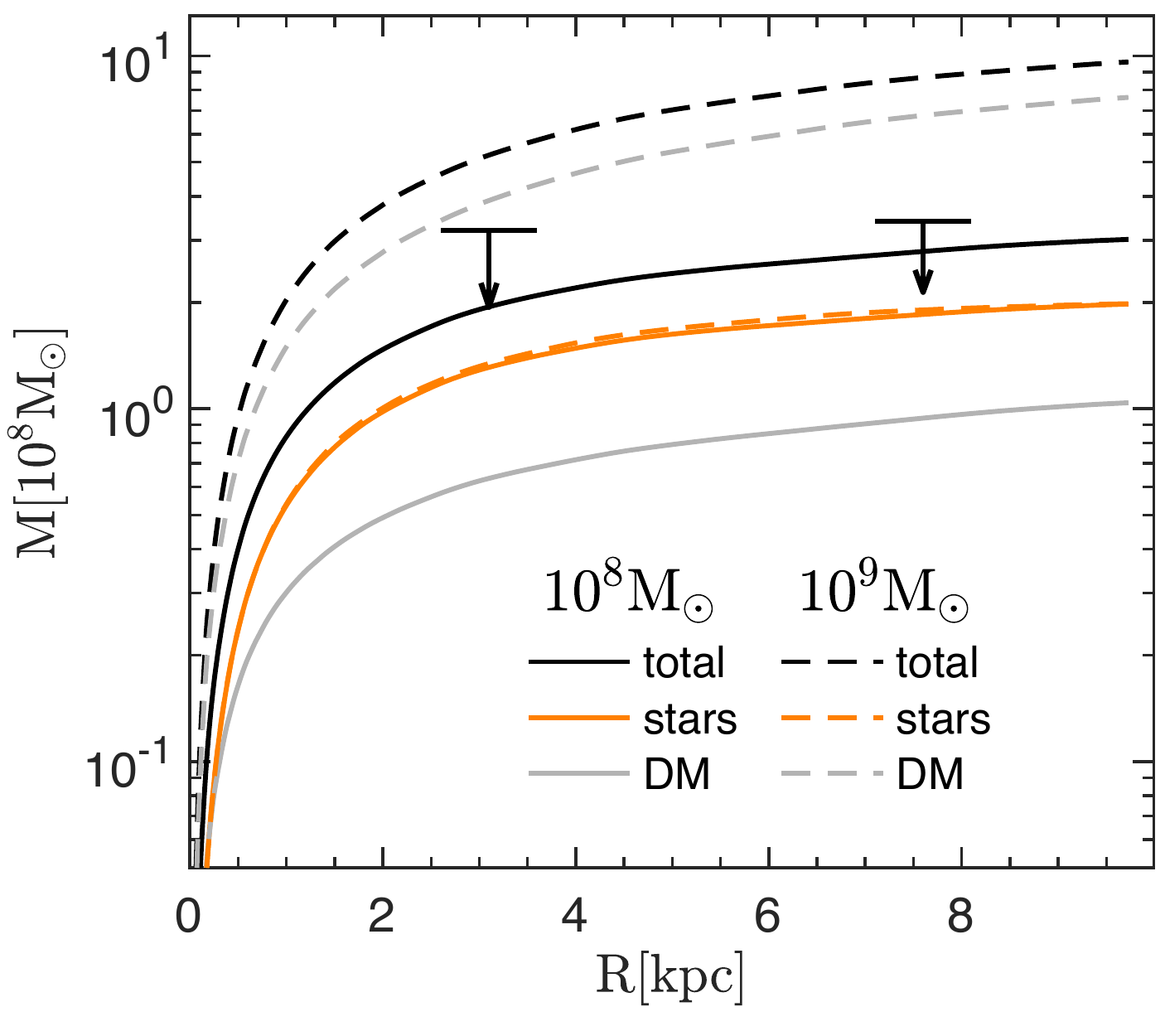}  % produced in Dropbox Technion ... Treecode/PltTwoSystem.m
  \vskip 0.0in
 \caption{ Mass profiles from the simulations run without GCs at an output time $t=1.5 \rm{Gyr}$. The lower curve 
represents a low mass galaxy close to 
 the observed  90\%  mass limits as indicated by the arrows. The upper curve is obtained from a  simulation with 
 4 times the mass in the lower curve.}
\label{fig:mprof}
\end{figure}

 The simulations are  run using the publicly available \textit{treecode} written by J. Barnes  \citep{Barnes1986},  with a force softening $\epsilon=0.05\, \textrm{kpc}$ and an opening angle criteria $\theta=1\textrm{rad}$. This value for $\epsilon$ is close to 
$b_\mathrm{min}\approx G M_\mathrm{GC}/V_\mathrm{GC}^2$, the impact parameter above  which 
encounters between the GC  and galactic particles are  important for dynamical friction  \citep{Binney2008}. Approximating the speed of GC by 
$V_\mathrm{GC}^2\approx GM_\mathrm{gal}/R_\mathrm{gal}$ we find $b_\mathrm{min}\approx R_\mathrm{gal} M_\mathrm{GC}/M_\mathrm{gal}\approx 0.1\, \mathrm{kpc}$ for $R_\mathrm{gal}\approx 10\, \mathrm{kpc}$ and $M_\mathrm{GC}/M_\mathrm{gal}\approx 0.01$. 
In any case, the uncertainty in  fixing  $\epsilon$ is of minor significance relative to the 
 the unknown detailed structure of the galaxy  and the actual three dimensional positions of the GCs in the observations.
Each simulation contains $2.4\times 10^5$  particles of equal mass, representing the galaxy without the GCs. 
A particle at position $r $ from the center, 
 is assigned a randomly oriented  initial velocity with a magnitude equal to the circular velocity, $V_c=\sqrt{GM(r)/r}$, where $M(<r)$ is the total (stars + DM) 
 within $r$. The initial configurations are evolved using the  \textit{treecode} for $1.5$ Gyr to obtain the corresponding  relaxed configurations.
 Fig.~\ref{fig:mprof}  shows the 2D mass profiles obtained from the simulation runs without GCs, for two halo masses, $10^8M_\odot$ and $10^9M_\odot$ as indicated in the figure. The evolved mass profiles  are  actually close to the respective profiles obtained from the initial conditions.
The stellar component in both simulation runs should match the observed stellar profile. Indeed, stellar distributions represented by the 
 orange (solid and dashed) curves for the low and high mass simulations,  match very well the 
corresponding profile in Fig.~4a in \cite{vanDokkum2018}. In computing the 2D DM profiles, we excise particles with (3D) distances larger than 10 kpc in the 
simulations. We obtain a good match with the DM mass profiles shown in the same figure of \cite{vanDokkum2018}.
 The arrows represent the $90\%$ 
confidence limits  on the mass estimates from the observations \citep{vanDokkum2018}.
The low  mass profile is close to the 90\% mass limits  from the observations. The circular velocity, $\sqrt{GM(r)/r}$,  for the lower mass profile
in the simulation, reaches a maximum of $\sim 17 \kms$ at $r=1-2\, \mathrm{kpc}$ and declines slowly at larger radii. 
The corresponding line of sight velocity dispersion is $\sim 9.8 \kms$, consistent with the observations. We have also checked (but do not show) that the simulated profiles  corresponding to a  distance $D=13 \, \mathrm{Mpc} $ of \NDF\ 
agree well with
 \cite{Trujillo2018}.
Once a relaxed  state is  reached, 
a massive particles representing a  GC is  placed in each simulation galaxy.
 For \cite{vanDokkum2018}
GCs of masses $1$, $2$, $3$ and $4$ million solar masses are placed at various distances  from the center with orbital
eccentricity of $e\sim 0.5$ \citep[e.g.][]{Benson2005,Wetzel2011}.
The observed \NDF\  is expected to be truncated at  $R_t \gtsim 7\,  \mathrm{kpc}$ by the tidal gravitational force field of  \textsc{NGC1052}, a much larger nearby elliptical galaxy 
at a projected distance of $\sim 100 \, \mathrm{kpc}$. Therefore, we only consider GC particles within $R_t$\footnote{Given a projected distance $R$ the probability for a 3D distance $r$ is $P(>r|R)\propto \ln (\sqrt{r^2/R^2-1}+r/R)$ for $R<R_t$ where $R_t$ is the truncation radius of the galaxy. This assumes the number density of GCs falls like $1/r$. }.

\begin{figure}
 \includegraphics[width=0.48\textwidth]{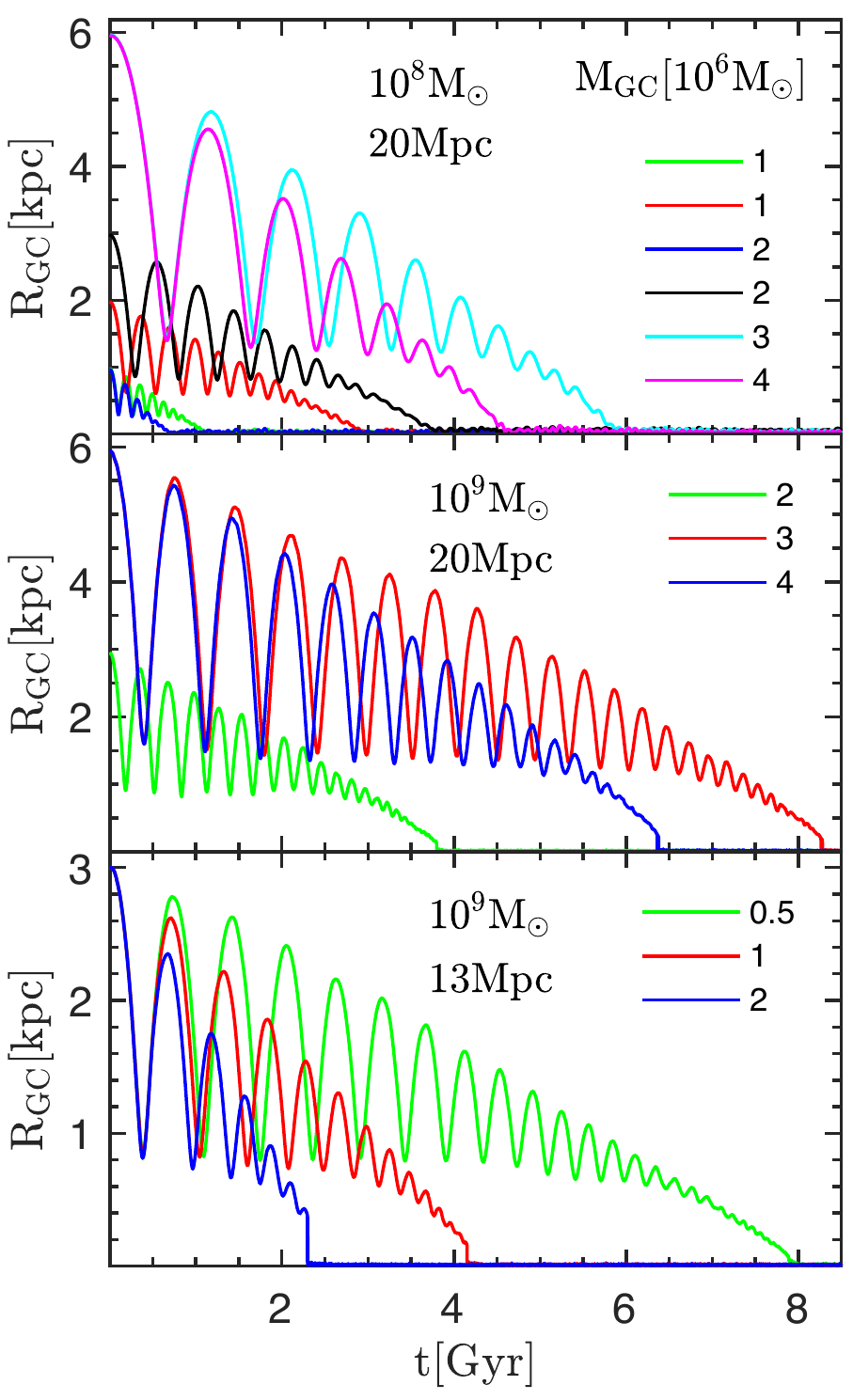}  % pltGChrsubplot1
  \vskip 0.0in
 \caption{{Top:} time dependence of the distance of GC particles  of various masses as indicated  on the top right hand corner of the figure. 
 The results correspond to  simulations assuming the galaxy embedded in a $10^8M_\odot$ DM halo mass at $D=20\, \mathrm{Mpc} $, as in \cite{vanDokkum2018}. {Middle:} same as in the top panel but for a $10^9 M_\odot$ DM halo.
{Bottom:} Results for a galaxy with a $10^9 M_\odot$ halo mass assumed at a $13\, \mathrm{Mpc}$ distance, as in \cite{Trujillo2018}.}
\label{fig:df}
\end{figure}

After the inclusion of GCs,  the simulations are run forward for $10\, \mathrm{Gyr}$, with an energy conservation to better than $1\%$.
For each simulated galaxy,  a  ``center"  is identified as the particle with the lowest potential energy.  Distances of 
GC particles are computed relative to the center in the corresponding simulation.
Fig.~\ref{fig:df} is a summary of the results for GCs distances versus time in the lower mass galaxy. The results are not reliable numerically for distances close to the softening parameter, $\epsilon$, e.g. at  a distance of 
$4\epsilon = 0.2 \mathrm{kpc}$  force bias introduced by the Plummer smoothing is about  10\%.
For the low mass galaxy run (top panel), the  orbit of the GC particle with $M_\mathrm{GC}=3\times 10^6 M_\odot$ (cyan line)  decays  within  $6\, \rm Gyr$ for an initial 
apocenter as large as $6\, \mathrm{kpc}$. Starting from an apocenter of $4\, \mathrm{kpc}$, the same curve shows that the orbit decays within $4\, \rm Gyr$. 
For an initial apocenter of $\ltsim 2 \,  \rm kpc$, orbits decay on a much shorter time scale of less than $3\rm Gyr$ even for the lightest particles.
At a  $20 \, \mathrm{Mpc}$ distance, the brightest  observed GC is with a mass in between the two largest GC particles and lies at a projected separation $\ltsim 2.4 \, \mathrm{kpc}$. The second brightest observed cluster is $\sim 2-2.5\times 10^6 M_\odot$ at a projected separation of $0.4\mathrm{kpc}$. The middle panel represents orbits in the simulations of a galaxy with a 10 times more  massive halo, still at a $20 \rm Mpc$  distance.
Orbital decay is clearly slower, however, it remains significant. Tracing the red curve, a GC with a $3\times 10^6 M_\odot$ starting at an apocenter of $4\, \rm kpc$, sinks to the center at $\ltsim 5\rm Gyr$. According to the green curve, a particles with $2\times 10^6 M_\odot$ (close to the mass of the second brightest observed GC), starting with apocenters 
of $3$ and $2\, \rm kpc$ reaches the center after $\sim 4\rm Gyr$ and $\sim 2\rm Gyr$, respectively. Note that the orbit of a  $2\times 10^6M_\odot$ particle 
starting from $3\, \rm kpc$  decays over  the same time scale in the low and high mass galaxies. This  is  seen by comparing the black curve in the top panel with the   green curve in the middle panel. 
Results for simulations corresponding to a galaxy at a $13\, \rm Mpc$  distance are represented in the bottom panel. 
On account of  the different assumed distances to \NDF , the GC masses in the simulations corresponding to  \cite{Trujillo2018}  are smaller than 
 for \cite{vanDokkum2018}. This boosts  the dynamical friction time scale but we must bear in mind that the lower distance implies smaller separations between the GCs and the center of \NDF .
 At a $13\, \rm Mpc$ distance, the mass of the brightest GC is $\approx 1.5\times 10^6 M_\odot$. Since  whole system is now less extended, we consider smaller separations than before. Starting from a $3\, \rm kpc$ a $2\times 10^6M_\odot$ particle reaches the center within $\ltsim 2.5\, \rm Gyr$, faster than in the top and middle panel. The reason is the smaller stellar mass in the $13 \, \rm Mpc$ distance galaxy which results in an overall less mass than the $20 \, \rm Mpc$ case, within the relevant radius. Therefore, also with the lower distance to \NDF , dynamical friction is expected tp play an important role.
 
\section{Discussion and Conclusions}
\label{sec:discussion}

We have argued that dynamical friction is an important process in \NDF\  for the mass range reported in \cite{vanDokkum2018}. 
Our findings imply  that the presence of the observed massive GCs at distances $\ltsim 6 \, \mathrm{kpc}$ 
is likely  inconsistent with the reported  mass estimate of \NDF .  
%As already argued above, according to Eq.~\ref{eq:DF}, $t_\mathrm{DF}\sim M_\mathrm{gal}^{1/2}$. This scaling is also supported by our simulations.
We have seen that boost of the galaxy mass by a factor  of 4 to $\gtsim 10^9 M_\odot $ 
is insufficient to suppress the effects of dynamical friction on the largest GCs in the relevant  range of distances from the center.
The scaling with $M_\mathrm{gal}$ implies that  boosting  the total mass of the galaxy by a factor of a  100, as would be expected for a typical galaxy  the same stellar mass, 
practically eliminates  the effects of dynamical friction. However, an order of magnitude 
increase over the reported mass limit in \cite{vanDokkum2018} is sufficient to increase to  $t_\mathrm{DF} \gtsim 10 \, \mathrm{Gyr}$. Thus, 
although our findings indicate that the ratio of DM to stellar mass in  \NDF\ should be on the order of a  few tens, they do not strictly 
require the typical high value of a few hundreds  expected from its  stellar mass.
A more detailed analysis of the mass constraints implied by the measured GC velocities might still 
yields that the dynamical mass might be large enough to avoid short dynamical friction timescales. 

Another way out is if the system is relatively young. But this possibility is yet to be demonstrated in a physical scenario that will yield consistency with all observations of the system  \cite{vanDokkum2018}. 
We have also run simulations where the whole system is subject to the external field of a nearby larger galaxy like NGC1052 in the vicinity of \NDF .
The results regarding orbital decay of GCs are unaltered by the inclusion of a (static)  external field, provided that they lie within the tidal radius of the simulated \NDF . It should be pointed out that the relative radial  velocity of \NDF\ relative to  NGC1052 is 
$293\kms$ \citep{vanDokkum2018}, while  line-of-sight velocity dispersion of the NGC1052 group is only $110\kms$  \citep{vanDokkum2018} which is consistent with the   circular velocity  of $200\kms$ measured from the \hi\ content of NGC1052  {Gorkom1986}.  Thus the relative speed of 
\NDF\ is close to the escape velocity from NGC1052.
At a projected distance of $100$ kpc between  the two galaxies, \NDF\   is likely to be just skimming past NGC1052. Nonetheless, even with tests where the external field generated from a galaxy with a 1D velocity dispersion of $300\kms$, the GCs either sank to the center or completely stripped
(for  large initial separations).

\cite{Ogiya2018} have shown that if i) \NDF\ is initially  harbored in a $4.9\times 10^{10}M_\odot$ halo with a large core and 
ii)  is on a highly radial orbits in the field of NGC1052, then gravitational tidal stripping 
could produce mass profiles consistent with the observations. As argued \cite{Ogiya2018} , this set-up is unlikely and systems like \NDF\ are expected to e rare. As mentioned above, we argue  the relative speed between \NDF\ and NGC1052 makes this set-up unrealistic. Thus, tidal stripping scenarios  \citep[e.g.][]{DiCintio2016,Carleton2018} are unlikely  to apply in the case of \NDF .

The situation is reminiscent of Fornax, the most massive satellite of the Milky Way. Fornax is the only satellite of the Galaxy containing   GCs; there are  five of them,   and they are observed at projected distances
$\sim 1 \, \mathrm{kpc} $ from its center. The dynamical friction time scale is short and at least the two  most massive of the GCs should have reached the center within a few $\rm Gyr$.    Fornax is DM dominated with a halo mass of $\sim 1.5 \times 10^8 M_\odot$, close to the mass of \NDF\  as given in \cite{vanDokkum2018}. But there are distinct differences between the two systems.  Fornax is much less spatially extended that the ultra diffuse \NDF\ 
if indeed at a distance of $D=13$ Mpc. Further, 
the  massive GC that Fornax  harbors is less that $0.3\%$ of its  mass. 
Therefore, none of our simulations corresponding to $20\, \rm Mpc$ can be directly associated with the relevant dynamical friction calculation done for Fornax.  At $D=13\, \rm Mpc$, the \NDF\ system becomes much more akin to Fornax. The larger GC mass in \NDF , could make  dynamical friction more problematic than in Fornax, but the tidal radius of Fornax is better determined and we have less freedom in fixing the 3D separation. 
Several ways out have been suggested to solve the Fornax mystery, such as having a core 
of constant density within $\sim 1 \, \mathrm{kpc} $ \citep{Goerdt2006}. This cored density profile is, however,   very hard to achieve within the context of viable warm DM models \citep{Maccio2012}. Further, even a mild cusp would would bring the orbits of some of the GCs to decay \citep{Cole2012}. 
 Other solutions same that the Fornax has only recently captured its GCs and that they are 
located near its tidal radius \citep{Cole2012,Oh2000,Angus2009}. This set-up  could be relevant for \NDF\ but it  seems unlikely and  its feasibility is hard to assess.

The orbit calculations done in this Letter  should serve as a general indication for the orbital decay times. 
The unknown detailed  structure of the \NDF\ and the availability of only partial phase space coordinates of the GC system, prevent an accurate determination of the orbits. 
Nonetheless, the  variety of numerical experiments presented here, sustained by analytic estimation, demonstrate clearly that typical decay time scales could  
comfortably shorter than the age of the system.  A clear conclusion from our work is that any mass model and formation scenario for \NDF\ should consider the 
constraints on GC orbital decay by dynamical friction. This statement is valid for the two distance measurements reported for this galaxy.
A full assessment of the parameter space of plausible mass profiles and orbital characteristics similar to  
the analysis \cite{Cole2012} for Fornax,  could be worthwhile. However, an extensive investigation is beyond the scope of this Letter.

\section*{Acknowledgements}

The author thanks Avishai Dekel, Fangzhou Jiang, Martin Feix and  Noam Soker for useful comments and Joshua Barnes for providing the  FORTRAN \textit{treecode} N-body code.
This research was supported by the I-CORE Program of the Planning and Budgeting Committee,
THE ISRAEL SCIENCE FOUNDATION (grants No. 1829/12 and No. 203/09)
and the Asher Space Research Institute.

% \bibliography{/Users/adi/Documents/Bibtex/library.bib}

\begin{thebibliography}{}
\expandafter\ifx\csname natexlab\endcsname\relax\def\natexlab#1{#1}\fi
\providecommand{\url}[1]{\href{#1}{#1}}
\providecommand{\dodoi}[1]{doi:~\href{http://doi.org/#1}{\nolinkurl{#1}}}
\providecommand{\doeprint}[1]{\href{http://ascl.net/#1}{\nolinkurl{http://ascl.net/#1}}}
\providecommand{\doarXiv}[1]{\href{https://arxiv.org/abs/#1}{\nolinkurl{https://arxiv.org/abs/#1}}}

\bibitem[{Angus \& Diaferio(2009)}]{Angus2009}
Angus, G.~W., \& Diaferio, A. 2009, MNRAS, 396, 887,
  \dodoi{10.1111/j.1365-2966.2009.14745.x}

\bibitem[{Arca-Sedda \& Capuzzo-Dolcetta(2016)}]{Arca2016}
Arca-Sedda, M., \& Capuzzo-Dolcetta, R. 2016, MNRAS, 464, 3060,
  \dodoi{10.1093/mnras/stw2483}

\bibitem[{Barnes \& Hut(1986)}]{Barnes1986}
Barnes, J., \& Hut, P. 1986, Nature, 324, 446, \dodoi{10.1038/324446a0}

\bibitem[{Behroozi {et~al.}(2010)Behroozi, Conroy, \& Wechsler}]{Behroozi2010}
Behroozi, P.~S., Conroy, C., \& Wechsler, R.~H. 2010, ApJ, 717, 379,
  \dodoi{10.1088/0004-637X/717/1/379}

\bibitem[{Benson(2005)}]{Benson2005}
Benson, A.~J. 2005, MNRAS, 358, 551, \dodoi{10.1111/j.1365-2966.2005.08788.x}

\bibitem[{Bettoni {et~al.}(2017)Bettoni, Nusser, Blas, \&
  Sibiryakov}]{Bettoni2017}
Bettoni, D., Nusser, A., Blas, D., \& Sibiryakov, S. 2017, JCAP, 2017,
  \dodoi{10.1088/1475-7516/2017/05/024}

\bibitem[{Binney \& Tremaine(2008)}]{Binney2008}
Binney, J., \& Tremaine, S. 2008, {Galactic dynamics} (Princeton University
  Press), 885.
\newblock \url{http://adsabs.harvard.edu/abs/2008gady.book.....B}

\bibitem[{Carleton {et~al.}(2018)Carleton, Errani, Cooper, Kaplinghat, \&
  Pe{\~{n}}arrubia}]{Carleton2018}
Carleton, T., Errani, R., Cooper, M., Kaplinghat, M., \& Pe{\~{n}}arrubia, J.
  2018, eprint arXiv:1805.06896.
\newblock \doarXiv{1805.06896}

\bibitem[{Chandrasekhar(1949)}]{Chandrasekhar1949}
Chandrasekhar, S. 1949, Rev. Mod. Phys., 21, 383,
  \dodoi{10.1103/RevModPhys.21.383}

\bibitem[{Cole {et~al.}(2012)Cole, Dehnen, Read, \& Wilkinson}]{Cole2012}
Cole, D.~R., Dehnen, W., Read, J.~I., \& Wilkinson, M.~I. 2012, MNRAS, 426,
  601, \dodoi{10.1111/j.1365-2966.2012.21885.x}

\bibitem[{{Di Cintio} {et~al.}(2017){Di Cintio}, Brook, Dutton, Macci{\`{o}},
  Obreja, \& Dekel}]{DiCintio2016}
{Di Cintio}, A., Brook, C.~B., Dutton, A.~A., {et~al.} 2017, Mon. Not. R.
  Astron. Soc. Lett., 466, L1, \dodoi{10.1093/mnrasl/slw210}

\bibitem[{D'Souza \& Rix(2013)}]{Souza2013}
D'Souza, R., \& Rix, H.~W. 2013, MNRAS, 429, 1887, \dodoi{10.1093/mnras/sts426}

\bibitem[{Einasto \& Haud(1989)}]{Einasto89}
Einasto, J., \& Haud, U. 1989, A\&A, 223, 89

\bibitem[{Famaey {et~al.}(2018)Famaey, McGaugh, \& Milgrom}]{Famaey2018}
Famaey, B., McGaugh, S., \& Milgrom, M. 2018, eprint arXiv:1804.04167.
\newblock \doarXiv{1804.04167}

\bibitem[{Frieman \& Gradwohl(1991)}]{Frieman1991}
Frieman, J.~A., \& Gradwohl, B.~A. 1991, PRL, 67, 1441,
  \dodoi{10.1126/science.260.5113.1441}

\bibitem[{Goerdt {et~al.}(2006)Goerdt, Moore, Read, Stadel, \&
  Zemp}]{Goerdt2006}
Goerdt, T., Moore, B., Read, J.~I., Stadel, J., \& Zemp, M. 2006, MNRAS, 368,
  1073, \dodoi{10.1111/j.1365-2966.2006.10182.x}

\bibitem[{Kesden \& Kamionkowski(2006)}]{Kesden2006}
Kesden, M., \& Kamionkowski, M. 2006, PRD, 74,
  \dodoi{10.1103/PhysRevD.74.083007}

\bibitem[{Keselman {et~al.}(2009)Keselman, Nusser, \& Peebles}]{Keselman2009}
Keselman, J.~A., Nusser, A., \& Peebles, P. J.~E. 2009, PRD, 80,
  \dodoi{10.1103/PhysRevD.80.063517}

\bibitem[{Lovell {et~al.}(2013)Lovell, Frenk, Eke, Jenkins, Gao, \&
  Theuns}]{Lovell2013}
Lovell, M.~R., Frenk, C.~S., Eke, V.~R., {et~al.} 2013, MNRAS, 439, 300,
  \dodoi{10.1093/mnras/stt2431}

\bibitem[{Ludlow {et~al.}(2013)Ludlow, Navarro, Boylan-Kolchin, Bett, Angulo,
  Li, White, Frenk, \& Springel}]{Ludlow2013}
Ludlow, A.~D., Navarro, J.~F., Boylan-Kolchin, M., {et~al.} 2013, MNRAS, 432,
  1103, \dodoi{10.1093/mnras/stt526}

\bibitem[{MacCi{\`{o}} {et~al.}(2012)MacCi{\`{o}}, Paduroiu, Anderhalden,
  Schneider, \& Moore}]{Maccio2012}
MacCi{\`{o}}, A.~V., Paduroiu, S., Anderhalden, D., Schneider, A., \& Moore, B.
  2012, MNRAS, 424, 1105, \dodoi{10.1111/j.1365-2966.2012.21284.x}

\bibitem[{Martin {et~al.}(2018)Martin, Collins, Longeard, \&
  Tollerud}]{Martin2018}
Martin, N.~F., Collins, M. L.~M., Longeard, N., \& Tollerud, E. 2018, ApJL,
  859, L5, \dodoi{10.3847/2041-8213/aac216}

\bibitem[{Moffat \& Toth(2018)}]{Moffat2018}
Moffat, J.~W., \& Toth, V.~T. 2018, eprint arXiv:1805.01117.
\newblock \doarXiv{1805.01117}

\bibitem[{Moster {et~al.}(2013)Moster, Naab, \& White}]{Moster2013}
Moster, B.~P., Naab, T., \& White, S. D.~M. 2013, MNRAS, 428, 3121,
  \dodoi{10.1093/mnras/sts261}

\bibitem[{Ogiya \& Go(2018)}]{Ogiya2018}
Ogiya, G., \& Go. 2018, eprint arXiv:1804.06421.
\newblock \doarXiv{1804.06421}

\bibitem[{Oh {et~al.}(2000)Oh, Lin, \& Richer}]{Oh2000}
Oh, K., Lin, D., \& Richer, H.~B. 2000, ApJ, 531, 727

\bibitem[{Pontzen \& Governato(2012)}]{Pontzen2012}
Pontzen, A., \& Governato, F. 2012, MNRAS, 421, 3464,
  \dodoi{10.1111/j.1365-2966.2012.20571.x}

\bibitem[{Tremaine \& Gunn(1979)}]{Tremaine1979}
Tremaine, S., \& Gunn, J.~E. 1979, PRL, 42, 407,
  \dodoi{10.1103/PhysRevLett.42.407}

\bibitem[{Trujillo {et~al.}(2018)Trujillo, Beasley, Borlaff, Carrasco, {Di
  Cintio}, Filho, Monelli, Montes, Roman, Ruiz-Lara, Almeida, Valls-Gabaud, \&
  Vazdekis}]{Trujillo2018}
Trujillo, I., Beasley, M.~A., Borlaff, A., {et~al.} 2018, eprint
  arXiv:1806.10141.
\newblock \doarXiv{1806.10141}

\bibitem[{van Dokkum {et~al.}(2018{\natexlab{a}})van Dokkum, Danieli, Cohen, \&
  Conroy}]{vanDokkum2018b}
van Dokkum, P., Danieli, S., Cohen, Y., \& Conroy, C. 2018{\natexlab{a}},
  eprint arXiv:1807.06025.
\newblock \doarXiv{1807.06025}

\bibitem[{van Dokkum {et~al.}(2018{\natexlab{b}})van Dokkum, Danieli, Cohen,
  Merritt, Romanowsky, Abraham, Brodie, Conroy, Lokhorst, Mowla, O'Sullivan, \&
  Zhang}]{vanDokkum2018}
van Dokkum, P., Danieli, S., Cohen, Y., {et~al.} 2018{\natexlab{b}}, Nature,
  555, 629, \dodoi{10.1038/nature25767}

\bibitem[{van Dokkum {et~al.}(2018{\natexlab{c}})van Dokkum, Cohen, Danieli,
  Kruijssen, Romanowsky, Merritt, Abraham, Brodie, Conroy, Lokhorst, Mowla,
  O'Sullivan, \& Zhang}]{vanDokkum2018a}
van Dokkum, P., Cohen, Y., Danieli, S., {et~al.} 2018{\natexlab{c}}, ApJL, 856,
  L30, \dodoi{10.3847/2041-8213/aab60b}

\bibitem[{Wetzel(2011)}]{Wetzel2011}
Wetzel, A.~R. 2011, MNRAS, 412, 49, \dodoi{10.1111/j.1365-2966.2010.17877.x}

\end{thebibliography}
% \end{document}

\end{document}